\documentclass[twocolumn,amsmath,amssymb,letterpaper,showpacs,prl]{revtex4}
\pdfoutput=1
\usepackage{graphicx}
\usepackage{bm}
\usepackage{dcolumn}

\begin{document}


\title{Particle motion during the compaction of granular matter}

\author{Steven Slotterback}
	\email{scsumd@umd.edu}
\author{Masahiro Toiya}
\author{Leonard Goff}
	\affiliation{Department of Physics, and IREAP, University of Maryland, College Park, Maryland, 20742}
\author{Jack F. Douglas}
\affiliation{Polymers Division and Center for Theoretical and Computational Materials Science, National Institute for Standards and Technology, Gaithersburg, Maryland}
\author{Wolfgang Losert}
	\email{wlosert@umd.edu}
\affiliation{Department of Physics, IPST and IREAP, University of Maryland, College Park, Maryland, 20742}

\date{\today}

\begin{abstract}
We track particle motions in a granular material subjected to compaction using a laser scattering based imaging method where compaction is achieved through thermal cycling.  Particle displacements in this jammed fluid correlate strongly with rearrangments of the Voronoi cells defining the local spatial partitioning about the particles, similar to previous observations of Rahman on cooled liquids.  Our observations provide further evidence of commonalities between particle dynamics in granular matter close to jamming and supercooled liquids.
\end{abstract}

\maketitle

Granular materials are strongly interacting particle systems that admit an admixture of collective particle motions, as in crystals, and random particle motions, as in simple fluids.  Under flow or deformation conditions these materials pass intermittently between a solid ``jammed'' state and a state of flow.  When a granular material jams, the individual particles are in stable mechanical equilibrium with their local neighbors \cite{PhysRevE.68.011306}.  Small perturbations such as tapping \cite{knight95} or shearing~\cite{losert04} can move the particles, and lead to the evolution from one jammed state to another, generally more compact state.  The increase in particle density with the number of perturbations is similar to the slow thermally driven dynamics of fluids close to their glass transition.  Observations of phenomena of this kind has led to the development of a general framework for the jamming transition, with temperature, density, and external forcing as the three main variables characterizing this dynamical transition ~\cite{PhysRevE.68.011306,LiuNagel}.

Although there have been a number of recent studies of particle trajectories in glass-forming colloidal fluids~\cite{PhysRevLett.89.095704,PhysRevE.60.5725} and other strongly interacting particle systems, little is known about the nature of particle trajectories in granular materials. 
While recent studies have successfully observed particle positions in a 3D jammed granular material~\cite{PhysRevE.68.020301,swinney07}, particle trajectories were not observed.
  It seems plausible that the particle motion has features in common with glass-forming liquids, such as the emergence of string-like collective motion \cite{douglas:144907}, and based on this hypothesis we examine whether the shape fluctuations in the Voronoi cells environments about the particles in compactified granular media correlate with the direction of particle displacements, as noticed by Rahman~\cite{Rahman66} in his pioneering studies of collective particle motions in cooled liquids.  On the other hand, the non-equilibrium nature of these {\it driven} particle systems leads us to expect differences from thermalized liquids that require a better understanding.  In this paper, we provide the first direct measurements of 3D particle trajectories during compaction, with a focus on the statistics of particle {\it motion}.  

In particular, we study particle rearrangements in a column of granular materials during the course of thermal cycling.  We use a laser sheet scanning method to reconstruct the configurations of the system at the end of each cycle.  The motion of individual particles is tracked over 10 cycles, and we analyze how the local geometry affects local particle dynamics.  

Thermal cycling provides a method of granular compaction without exciting strong motions of particles. 
By thermal cycling we meanalternately heating and cooling the material and container, with one heating and cooling cycle corresponding to one perturbation.  This technique leads to exponential compaction of the material with two characteristic timescales~\cite{schiffer06} if the granular material and container have different thermal expansion coefficients.  While most methods of forcing rearrangements of granular matter, such as tapping of the boundaries or air fluidization agitate individual particles significantly, i.e. provide particles kinetic energy relative to their neighbors, thermal cycling provides a quasi-stationary method of evolving the structure towards a jammed state.
 
{\bf Experimental Setup:}
The granular material used is soda-lime glass beads (diameter $D=3{\rm\:mm}\:\pm\:0.3{\rm\:mm}$), poured into a transparent polymethylpentene (PMP) cylinder of diameter 5cm (16.6 D) to a height of $19.5{\rm\:cm}\:\pm\:0.5{\rm\:cm}$ (63.3 D to 66.6 D).  This is the same materials as used in previous work on thermal cycling~\cite{schiffer06}.  The thermal expansion coefficient for PMP is $1.17 \times 10^{-4} {\rm\:K^{-1}}$, about one order of magnitude larger than that of the beads ($9 \times 10^{-6} {\rm\:K^{-1}}$) ~\cite{schiffer06}.  The beads were immersed in index-matching oil (Cargille Labs Type DF~\footnote{Identification of a commercial product is made only to facilitate reproducibility and to adequately describe procedure. In no case does it imply endorsement by NIST or imply that it is necessarily the best product for the procedure.}) which contains $2.08\:\mu{\rm g/mL}$ laser dye (Nile Blue 690 perchlorate).  A transparent weight of $145.5{\rm\:g}$ was placed on top of the particles to monitor the filling height and to apply a controlled vertical force.

The cylinder is placed inside a transparent water tank for thermal cycling at a rate of $16^{o}$ C per hour.  As the PMP container was heated/cooled the cross-sectional area expanded/contracted by $0.46 {\rm\:mm^2}$ for every degree heated/cooled ($6.5\:\%$ of a single particle cross-section).

While bead expansion was negligible, the sample and oil are index-matched at room temperature.  Since the index of refraction changes with temperature differently for glass and oil, we can only image our sample at the end of each thermal cycle.

\begin{figure}[t,h]
\centering
\includegraphics[width=0.48\textwidth]{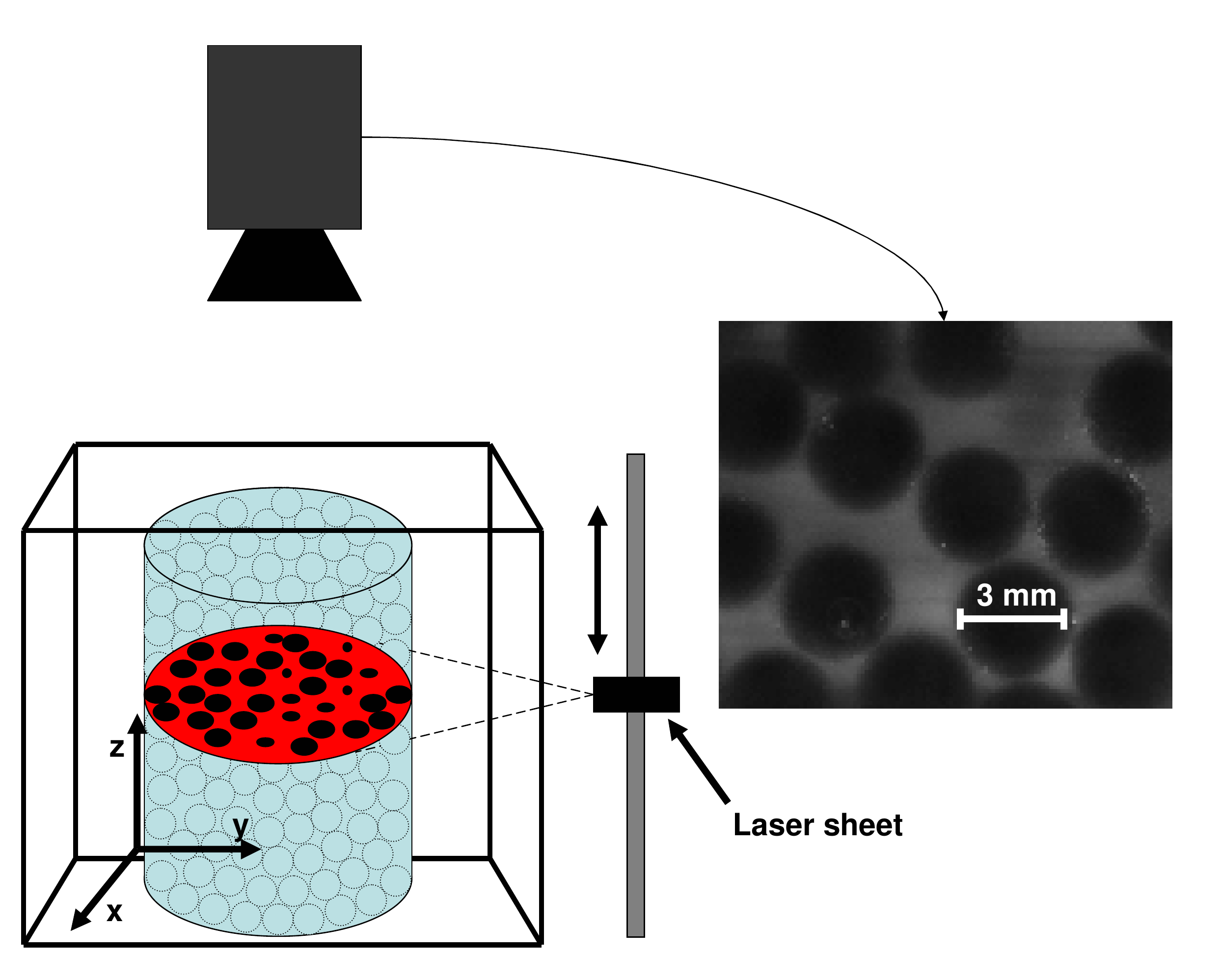}
\caption{Experimental setup:  Glass beads immersed in index matching oil in a cylinder are subjected to thermal cycling via a water bath.  The fluorescently dyed fluid is imaged via a laser sheet that is moved across the sample.  One of 600 cross-sectional images shown.}
\label{fig:setup}
\end{figure}

Three dimensional images of the system at each cycle were obtained using a laser sheet scanning method, first described in~\cite{Losert_pene}.  Briefly, a laser sheet is sent through the specimen at the excitation frequency of the laser dye to produce a cross-sectional image of the granular material in which the particles appear as dark circles within the dyed fluid (see Fig.~\ref{fig:setup}).  Images are taken with a high sensitivity cooled CCD camera (Sensicam, PCO).  To obtain a three dimensional image, the laser is translated in $100\:\mu{\rm m}$ increments along the axis of the cylinder for a distance of 60 mm (20 D), starting from the top of the pile.  The camera is not translated, but the changing object distance is accounted for in the analysis.
    
A 3D bandpass filter is used to reduce noise and to smooth out images of the beads.  A particle tracking algorithm by Crocker~\cite{crocker96} and Weeks~\cite{weeks07} detects the centers of $> 98\:\%$ of all beads to within $100\:\mu {\rm m}$, which is smaller than the uncertainty in the bead diameters.  The remaining $< 2\:\%$ of bead positions are detected manually with comparable resolution using an interactive 3D visualization of images with an overlay of extracted particles.  This allows us to generate complete maps of 3D jammed states.  Particle rearrangmenets are followed through multiple thermal cycles.  The particles move by less than 0.5 D in each thermal cycle, a scale small enough to allow for reliable particle tracking.

\begin{figure}[t,h]
\centering
\includegraphics[width=0.48\textwidth]{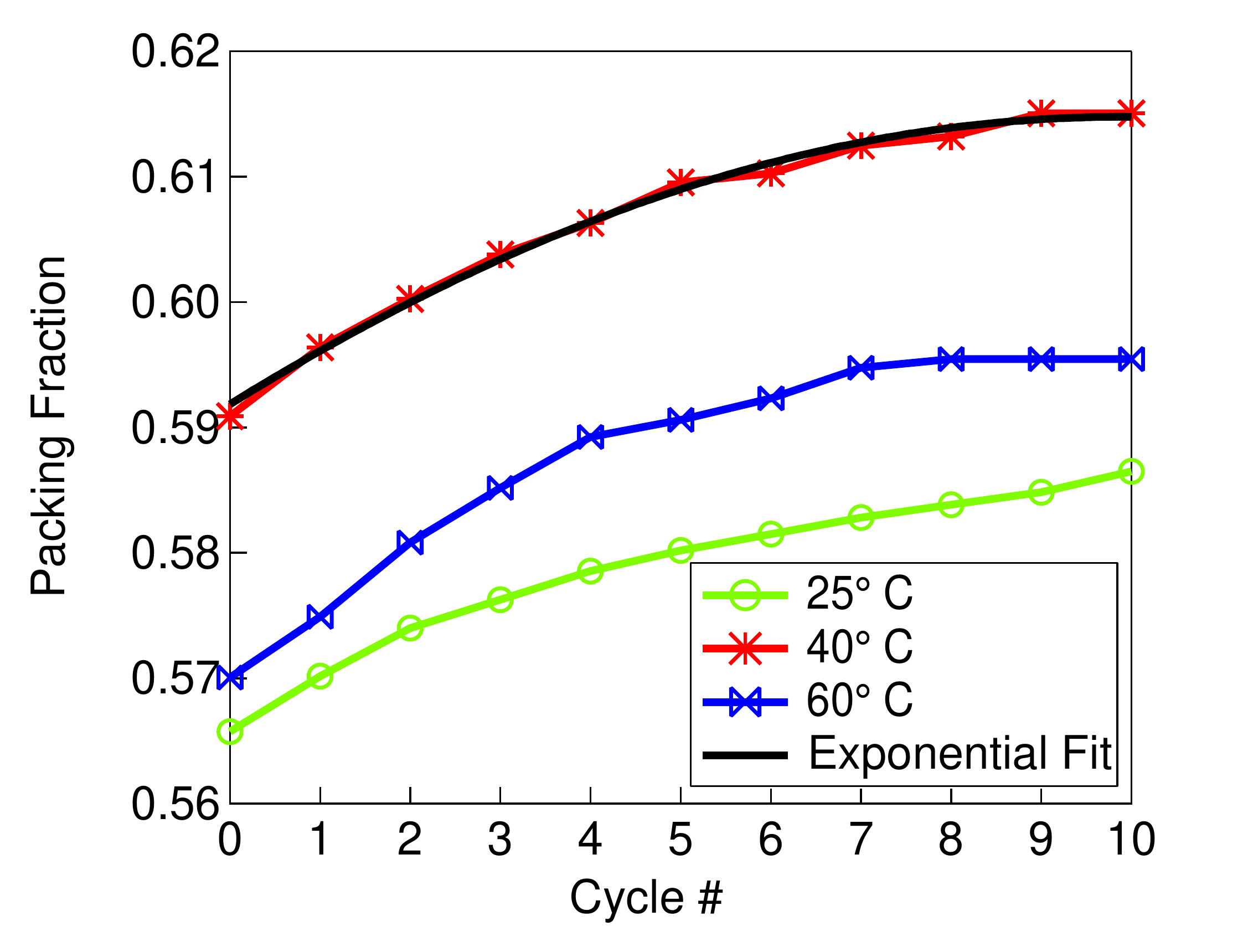}
\caption{Packing fraction vs. cycle number for various temperature differentials}
\label{fig:density}
\end{figure}

{\bf Results:}
We apply ten thermal cycles for three different temperature differentials: $25{\rm^{o}\:C}$, $40{\rm^{o}\:C}$, and $60{\rm^{o}\:C}$.  The average packing fraction is determined from the vertical position of the flat glass weight on top of the pile in the 3D image.  The initial conditions are generated by letting particles settle into the oil filled cylinder.  After the particles settle, the glass top is allowed to sink onto the surface of the fluid immersed particles.  Filling in oil first is necessary to eliminate air bubbles from the sample, but leads to significant variability in the initial packing fraction.  Nevertheless, the increase in packing fraction during thermal cycling can be fit with an exponential $\rho(t)=\rho_0 -A\:e^{t/\tau{}}$ as shown in Fig.~\ref{fig:density} with $\rho_0 = 0.619 \pm 0.0008$, $A = 0.028 \pm 0.0007$, and $\tau=4.87\:{\rm cycles} \pm 0.31\:{\rm cycles}$.  The compaction rate is slower than the faster of the two time scales, 2.72 cycles, reported by~\cite{schiffer06}, consistent with a weaker forcing of the particles due to buoyancy forces.  

\begin{figure}[t,h]
\centering
\includegraphics[width=0.48\textwidth]{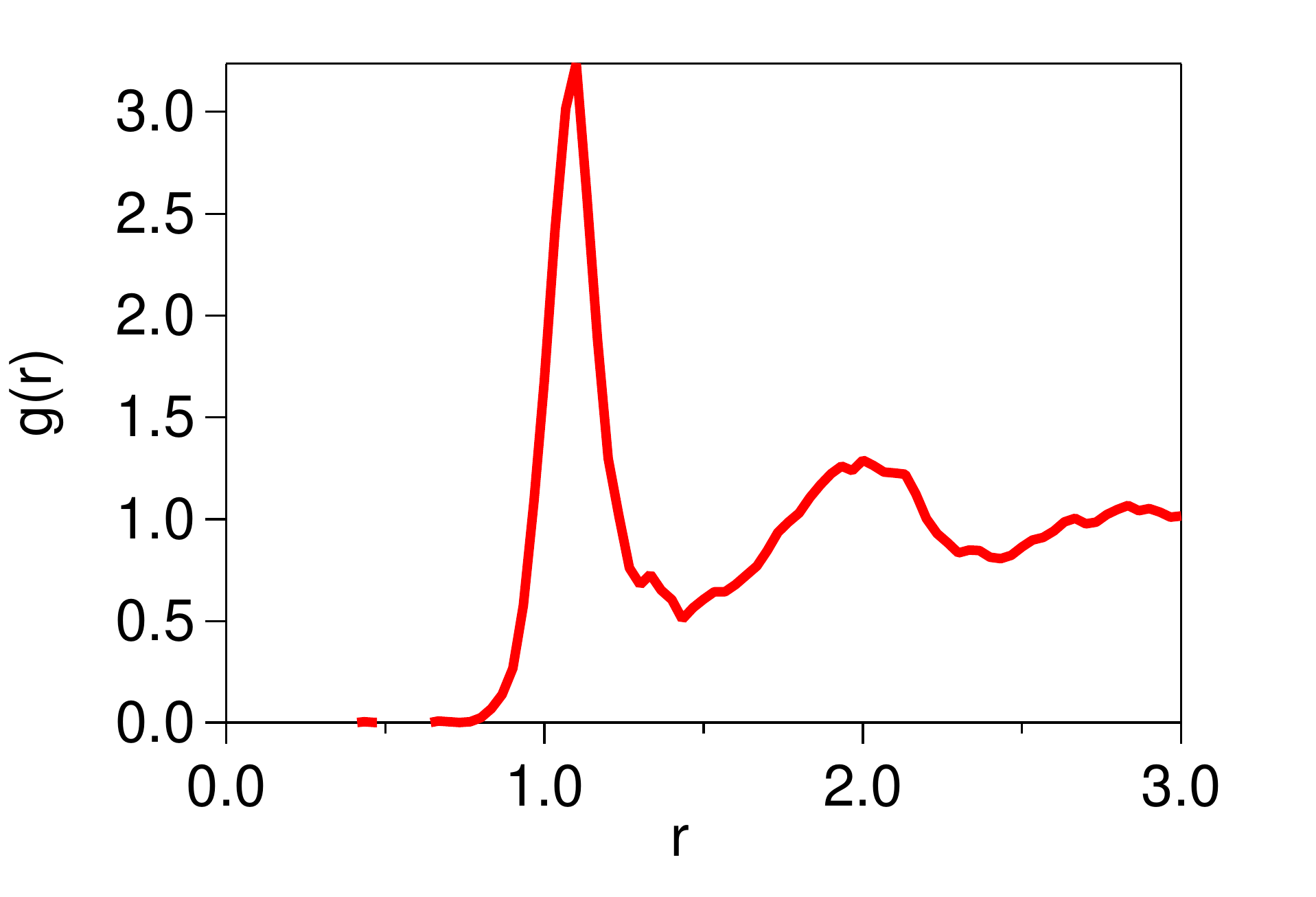}
\caption{Pair correlation function $g(r)$ averaged over 10 thermal cycles.}
\label{fig:GofR}
\end{figure}

We find that the pair correlation function $g(r)$ does not evolve significantly during compaction, as shown in Fig.~\ref{fig:GofR}.  This is not surprising since the particles only need to move closer by about $0.001\:D$ on average to achieve an increase in packing fraction of $1.5\:\%$.  The overall shape of $g(r)$ is consistent with a random arrangement of particles in the presence of some ordering at the boundaries.  

\begin{figure}
\centering
\includegraphics[width=0.48\textwidth]{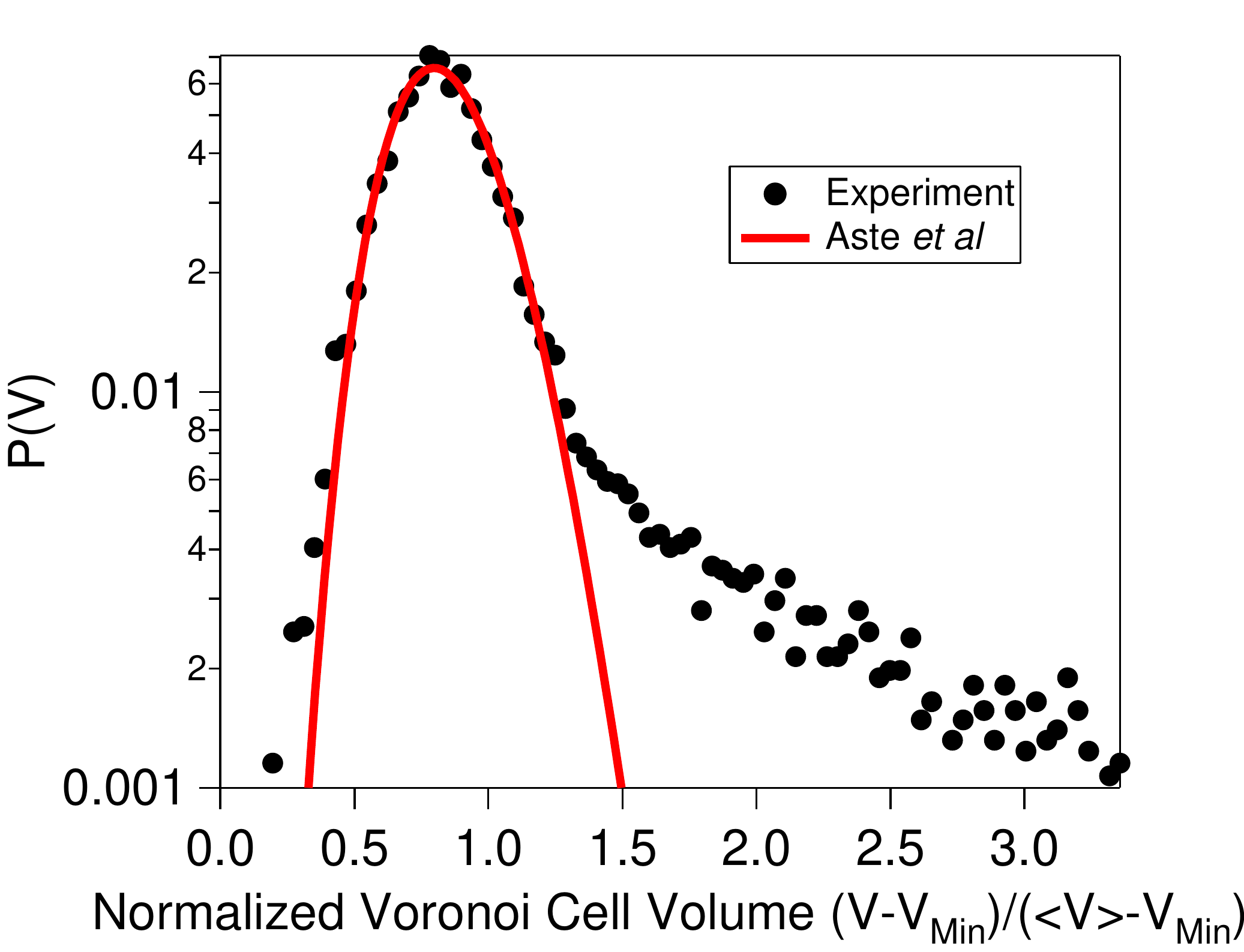}
\caption{Histogram of Voronoi volumes}
\label{fig:Voronoi}
\end{figure}

The Voronoi construction defines a unique partitioning of space where each particle is assigned a local neighborhood, the Voronoi cell, corresponding to the region of space closest to a given particle.  The Voronoi cell structures were determined using the QHULL algorithm~\cite{235821}.  To avoid boundary artifacts, Voronoi cells were computed for all particles, but only the volumes of particles that are not on the boundaries we used in the analysis.  The distributions of Voronoi volumes is shown in Fig. ~\ref{fig:Voronoi}.  The distribution is non-gaussian with a sharp cutoff at small volumes and a broad range of large volumes, similar to distributions observed in simulations in glass-forming liquids ~\cite{PhysRevLett.89.125501} and in recent experiments of granular materials ~\cite{swinney07}.  
One possible functional form of the distribution of voronoi volumes proposed by Aste 
{\it et al.}~\cite{swinney07} contains only one fitting parameter $k$.  Fitting our Voronoi cell volume distributions to this function yields $k=24$, appreciably larger than the $k$ value, $k=12$, suggested by Aste {\it et al.}~\cite{swinney07}, however. The larger polydispersity in bead diameters in the experiments presented here ($10\:\%$ as opposed to $2\:\%\:{\rm to}\:5\:\%$), could explain the wider distribution of Voronoi volumes.  We find that the distribution of Voronoi volumes does not change significantly as the system is compacted.

\begin{figure}
\centering
\includegraphics[width=0.38\textwidth]{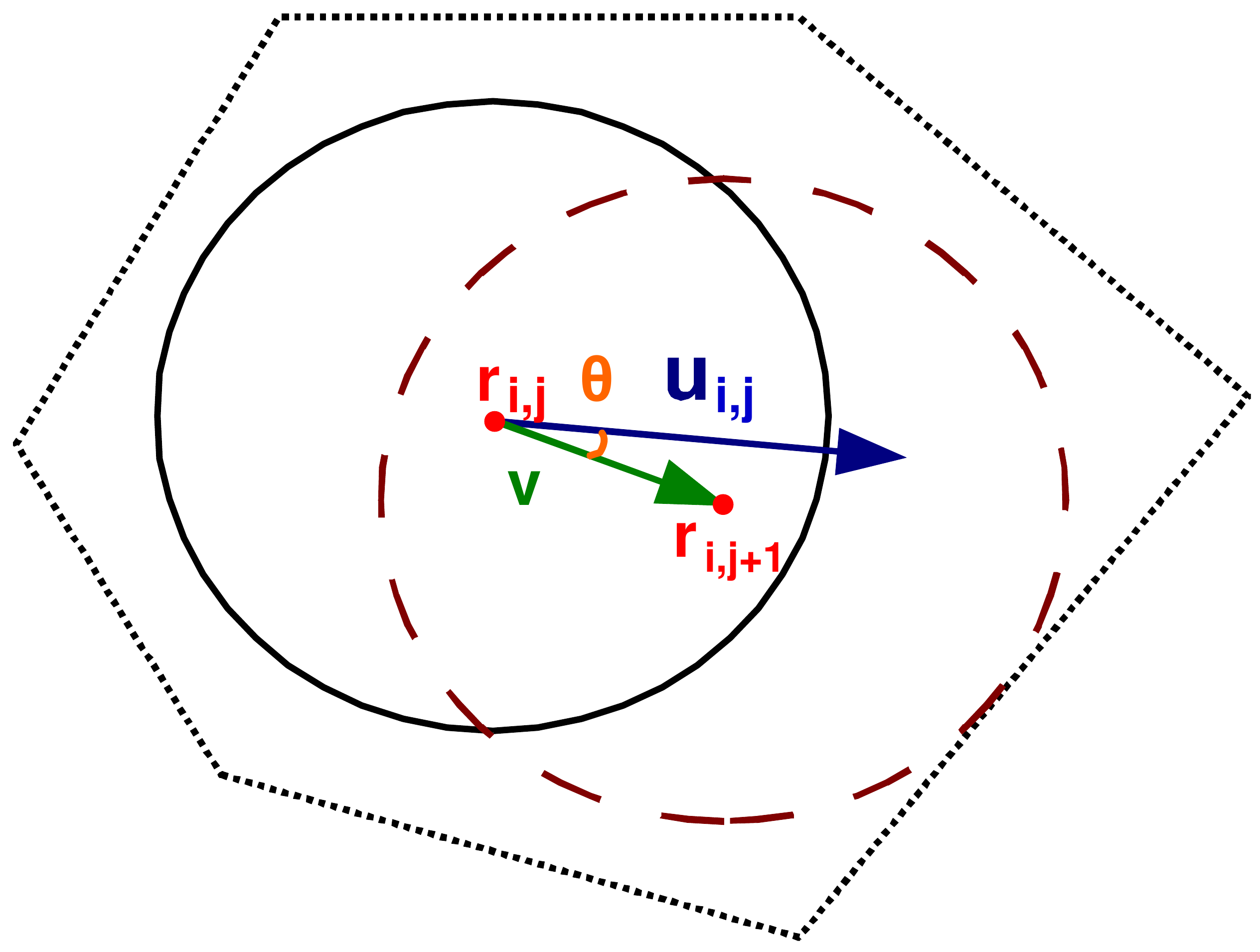}
\caption{Definition of particle displacement vector {\it v} and {\it u} vector 
for a cross-section of the 3D Voronoi volume.  Since this is a 
cross-section of the 3D Voronoi cell, the particle need not be in 
contact with any of the edges of the Voronoi volume.}
\label{fig:uvec}
\end{figure}

Following Rahman's investigation of cooled liquids, we analyze particle motion and its relation to the Voronoi cell shape.  We define the vector $\vec{v}_{i,j}$ to be the displacement of the center of particle $i$ from cycle $j$ to $j+1$ (see Fig.~\ref{fig:uvec}).  Next, we define a vector describing the Voronoi cell shape relative to the particle center.  From the perspective of the particle, the vertices of the vornonoi construction correspond to the direction in which a particle ``sees'' a void between three neighbors.  The mean position of all vertices therefore indicates where a particle would see
more local void spaces.  The vector $\vec{u}_{i,j}$ defines the displacement from the center of particle $i$ at cycle $j$ to the mean position of its corresponding Voronoi vertices, as shown in Fig.~\ref{fig:uvec}.  The averages for $|\vec{u}|$ and $|\vec{v}|$ are approximately 1 mm (0.33 D) larger than the accuracy with which we detect particle positions.

\begin{figure}[b]
\centering
\includegraphics[width=0.48\textwidth]{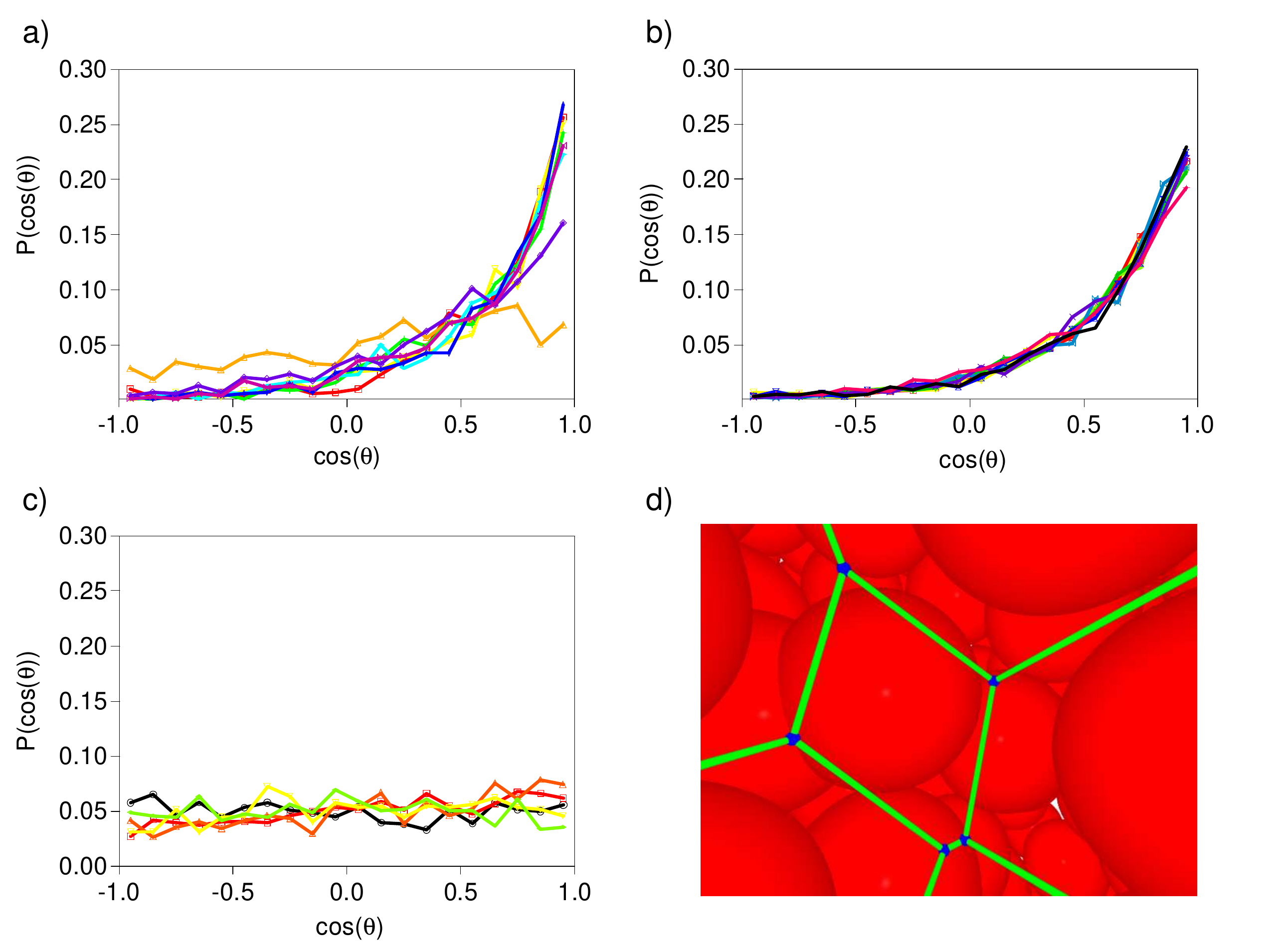}
\caption{{\textbf a)} Preferred Direction PDFs for system with $\Delta T = 40{\rm^{o}\:C}$ for each thermal cycle. {\textbf b)} Preferred Direction PDFs for system with $\Delta T = 25{\rm^{o}\:C}$ for each thermal cycle. {\textbf c)} Preferred Direction PDFs for system at constant temperature {\textbf d)} 3D reconstruction of system from the perspective of a given particle facing five vertices of its Voronoi cell.  Notice that the vertex is located in the void between particles.}
\label{fig:udotv}
\end{figure}

To determine the correlation between particle displacement and Voronoi shapes, we then consider the alignment of $\vec{u}_{i,j}$ and $\vec{v}_{i,j}$.  Specifically, we calculate $\cos(\theta)$ for each image, where $\theta$ is the angle between $\vec{u}_{i,j}$ and $\vec{v}_{i,j}$ for each particle.  The probability distribution of $\cos{\theta}$ averaged over all cycles is shown in Fig.~\ref{fig:udotv}.  In addition, we show the distribution for the first and the last cycle, and a control for two 3-D images without thermal cycling between the images.  We find that the distribution peaks at $\cos{\theta} = 1$, indicating that the particle moves toward the center of the vertices of its Voronoi cell.  Particle motion and Voronoi shape are indeed correlated.

{\bf Discussion:}

By using the laser sheet scanning method, we visualize the internal structure and dynamics of a jammed granular material.  Under quasi-static forcing via thermal cycling flows are slow enough (less than $\approx 0.5\:{\rm D}$ between frames), so we can observe the microscopic dynamics of this strongly interacting particle system.  While the differences in local structures are too small to be observable in local distributions such as $g(r)$ or the Voronoi volume distribution, the Voronoi volume appears to hold important clues for the future dynamical evolution of the particles.  Voronoi reconstruction indicates that particles tend to move toward the centers of Voronoi vertices as long ago observed in Rahman's pioneering simulations of particle motion in cooled liquids ~\cite{Rahman66}.  This correlated displacement in our granular fluid is independent of the size of the Voronoi volume, the velocity of particles, and the position of particles within the granular column.

Rahman~\cite{Rahman66} and others following him ~\cite{zwanzig:295} inferred from the correlations in the tracer particle displacement and the Voronoi cell shape that the particles were moving in tube-like environments, thus anticipating the string-like motion in cooled liquids that has recently become widely appreciated~\cite{PhysRevLett.80.2338,douglas:144907}.  This raises questions about whether such collective interparticle displacements also characterize particle motion in compactified granular materials.  The absence of equilibrium in granular fluids, however, complicates the definition of ``mobile'' and collectively moving particles in granular materials, since there is no model such as Brownian motion to provide a ``baseline'' for this comparison.  A recent study of a quasi-2D air-driven granular material ~\cite{Keys07} indicates the presence of collective string-like motion as in cooled liquids so the general presence of such collective motions in driven granular media seem highly plausible.  However, we expect important differences to emerge in systems such as ours where the volume of the material is evolving in time with compaction.  We plan to study the evolutionary or aging dynamics of our granular fluid in the future. 
\begin{acknowledgments}
We thank Krisztian Ronaszegi for assistance in data analysis.  This work was supported by NSF grants CTS0457431 and CTS0625890.  Official contribution of the National Institute of Standards and Technology - Not subject to copyright in the United States.
\end{acknowledgments}
	\bibliography{ThermoCycle_02_02_08}

\end{document}